\documentclass [12pt,twoside]{article}           
\usepackage[left,modulo]{lineno}
\usepackage{setspace}

\countdef\arxiv=201

\ifnum\arxiv=0 \input cpreb.tex \fi

\ifnum\arxiv=1

\usepackage{epsfig,times,lscape}
\usepackage[usenames]{color}

    \ifnum\count102=1

    \fi

    \ifnum\count102=2
\fi

\newcommand{\nc}{\newcommand}

     \ifnum\count101=1
\nc{\qI}[1]{\section{{#1}}}
\nc{\qA}[1]{\subsection{{#1}}}
\nc{\qun}[1]{\subsubsection{{#1}}}
\nc{\qa}[1]{\paragraph{{#1}}}

\def\qpar{\vskip 2mm plus 0.2mm minus 0.2mm}
\def\qL{\hfill \break}
     \fi 

      \ifnum\count101=2
 \nc{\qI}[1]{\parindent=0mm \vskip 8mm 
{\centerline{\LARGE \color{red}#1}}\vskip 3mm}
\nc{\qA}[1]{\vskip 2.5mm \noindent 
{{\bf\large\color{blue}  #1}} \vskip 1mm \parindent=0mm}
 \nc{\qun}[1]{\vskip 1mm \noindent {\sl #1 }\quad }

\def\qL{\hfill \break}
\def\qpar{\vskip 2mm plus 0.2mm minus 0.2mm}

      \fi

\nc{\qfoot}[1]{\footnote{{#1}}}

      \ifnum\count101=1
\def\qbu{\hfill \par \hskip 6mm $ \bullet $ \hskip 2mm}
\def\qee#1{\hfill \par \hskip 6mm (#1) \hskip 2 mm}
      \fi
      \ifnum\count101=2
\def\qbu{\hfill \par \hskip 4mm $ \bullet $ \hskip 2mm}
\def\qee#1{\hfill \par \hskip 4mm (#1) \hskip 2 mm}
      \fi

\def\qparr{ \vskip 1.0mm plus 0.2mm minus 0.2mm \hangindent=10mm
\hangafter=1}

     \ifnum\count101=1 
 
     \fi
     \ifnum\count101=2

  \def\qcitb#1{\noindent \hbox to 102mm{\hfill \small #1} \vskip 1mm}
      \fi

 \def\qpages#1{\count102=0{\loop\advance\count102 by 1
 \null \vfill\eject \ifnum\count102<#1 \repeat}}

\def\qv{\vskip 0.1mm plus 0.05mm minus 0.05mm}
\def\qhu{\hskip 0.6mm}
\def\qhv{\hskip 3mm}

\def\qhw{\hskip 1.5mm}
\def\qleg#1#2#3{\noindent {\bf \small #1\qhw}{\small #2\qhw}{\it \small #3}\qv }
\fi

\begin{document}
\thispagestyle{empty}



\markboth{{\sl \hfill  \hfill \protect\phantom{3}}}
        {{\protect\phantom{3}\sl \hfill  \hfill}}

\color{yellow} 
\hrule height 20mm depth 10mm width 170mm 
\color{black}
\vskip -1.8cm 

\centerline{\bf \Large Phenomenology of infant death rates}
\vskip 2mm
\centerline{\bf \Large Identification of the peaks of viral and 
bacterial diseases}
\vskip 2mm
\centerline{\bf \Large }
\vskip 15mm
\centerline{\large 
Peter Richmond$ ^1 $ and Bertrand M. Roehner$ ^2 $
}

\vskip 4mm
\large

{\bf Abstract}\quad 
After birth setting up an effective immune system is a major
challenge for all living organisms. 
In this paper we show
that this process can be explored by using the 
age-specific infant death rate as a kind of sensor.
This is made possible because, as shown by the authors
in Berrut et al. (2016), between birth and a critical
age $ t_c $, for all mammals the death rate decreases
with age as an hyperbolic function. 
For humans $ t_c $ is equal
to 10 years. At some ages the hyperbolic fall displays 
spikes which, it is assumed, correspond to specific events
in the organism's response to exogenous factors.
One of these spikes occurs 10 days after birth and there is
another at the age of 300 days. It is shown that
the first spike is related to viral infections
whereas the second is related to bacterial diseases.
By going back to former time periods during which infant
mortality was much higher than currently,
it is possible to get a magnified view of these peaks
which in turn may give us useful information about
how an organism adapts to new conditions.
Apart from pathogens,
the same methodology can be used to study the response
to changes in other external conditions, e.g.
supply of food, temperature or oxygen level.

\vskip 2mm
\centerline{\it \small Provisional. Version of 5 October 2016. 
Comments are welcome.}
\vskip 2mm

{\small Key-words: death rate, infant mortality, viral, bacterial,
diseases, spectroscopy}

\vskip 2mm

{\normalsize 
1: School of Physics, Trinity College Dublin, Ireland.
Email: peter\_richmond@ymail.com \qL
2: Institute for Theoretical and High Energy Physics (LPTHE),
University Pierre and Marie Curie, Paris, France. 
Email: roehner@lpthe.jussieu.fr
}

\vfill\eject

\qI{Introduction}

Why did we entitle this note ``Phenomenology of death rates''?
In physics the word ``phenomenology'' refers to
the interface between experimental results and theory.
Its purpose is to bring together appropriate data and organize 
them in the best way for testing any existing predictions.
When no theory is
available the objective is to bring to light phenomenological
regularities and to recommend further experiments to probe
their validity.
\qpar
In the social sciences such a program usually cannot
be carried out because of limitations in the data 
available and the impossibility to produce new ones to fill
the gaps.
\qpar
Age-specific death rates by cause of death are probably
the variables which come closest to the situation prevailing for
physical data. There are two main reasons for that.
\qbu Death records were the first demographic data to be 
collected in a systematic way. In the mid-19th century
in all industrialized countries
organizations were set up in order to record the age, 
date and cause of deaths. Recording first started
in cities and was progressively extended to rural areas.
In the case of the United States, the so-called ``death registration
area'' covered about 40\% of the population in 1900 and
96\% in 1930.
\qbu By the end of the 19th century medical authorities and
statisticians from over all developed countries agreed on an
international classification of the causes of death.
The adoption of this classification was an essential step
because it ensured cross country comparability.
\qbu

The age-specific postnatal death rate follows
a hyperbolic decrease $ \mu_b(t)\sim 1/t^{\gamma} $ 
until a critical age $ t_c $ after
which it starts to increase; then, for most species, it
keeps increasing until death. For humans, $ t_c= 10 $ years.
This pattern holds, not only  for the global (i.e. ``all causes'')
mortality but also for individual diseases, albeit with distinct
values of $ \gamma $. Fig. 1 illustrates this statement for
the case of metabolic diseases. 
%
\begin{figure}[htb]
\centerline{\psfig{width=14cm,figure=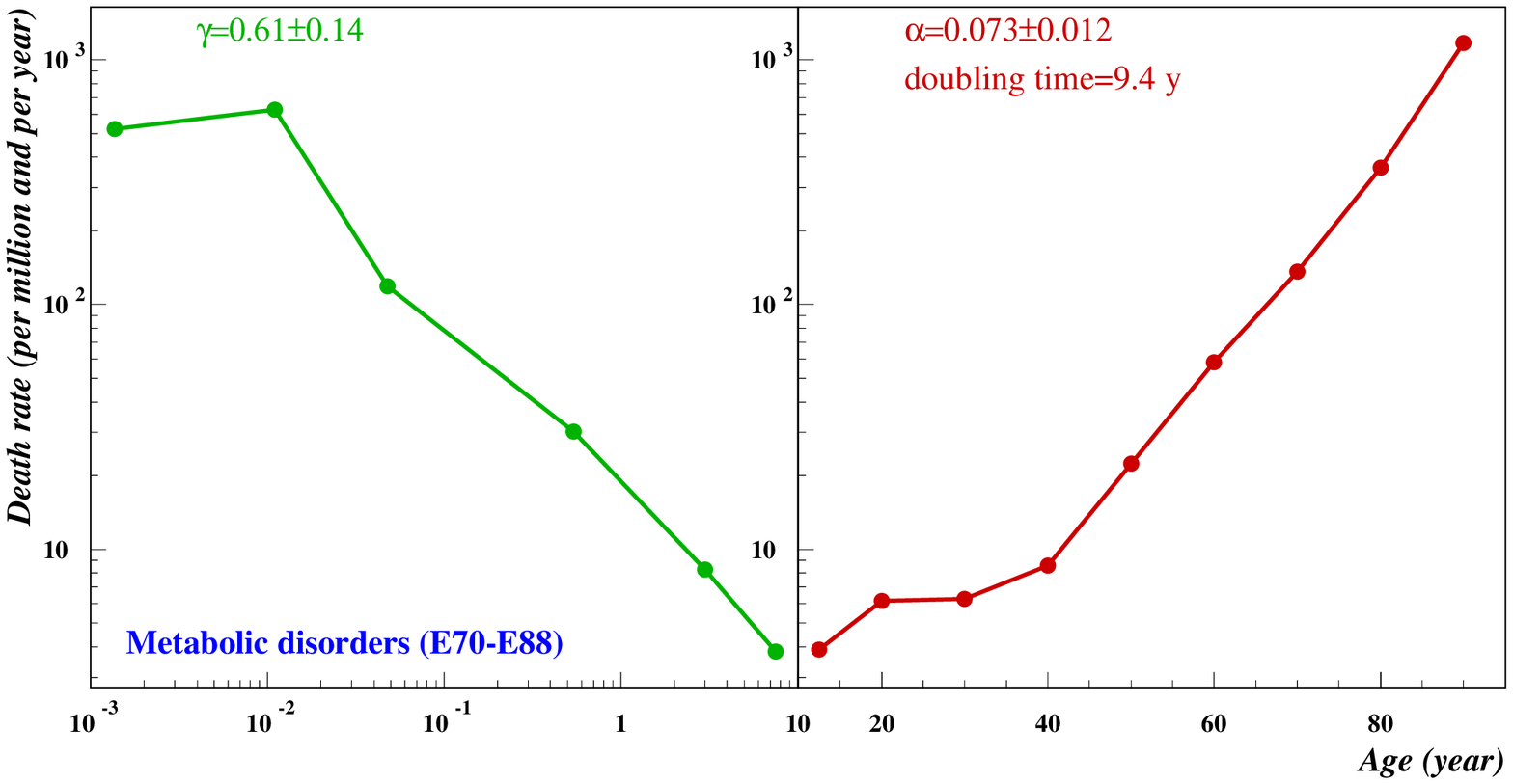}}
\qleg{Fig.\qhu 1\qhv Death rate due to metabolic disorders,
USA 1999-1984.}
{Included  in this class of disorders are for instance
intolerance to lactose and defects in the metabolism of
carbonhydrate. These down- and up-going curves 
($ \mu_b(t)\sim 1/t^{\gamma}, \mu(t)\sim \exp(-\alpha t) $
respectively) are 
similar to those for ``all causes'' death rates except that
the exponents are not the same. For $ \gamma $ it is
$ 0.99 $ instead of $ 0.61 $ here. For 
the doubling time in the
adult phase it is $ 8.6 $ years instead of $ 9.4 $ years here.
$ \gamma $ and the doubling time can be seen as signatures
of a specific class of diseases.} 
{Source:  CDC ``Wonder'' database 1999-2014
(http://wonder.cdc.gov/ucd-icd10.html).
The data are average annual death rates over the 16-year
long interval 1999-2014.}
\end{figure}
Why did we illustrate this pattern through metabolic
disorders instead of a case that would be
closer to the class of infectious
diseases which will be considered later on in this paper?
We wanted to emphasize that the strengthening (or
weakening) of the immune system was not directly
responsible for the
hyperbolic fall (nor for the exponential growth). This does
of course not mean that they do not affect death rates
in any way but only that the basic pattern is due to
another mechanism. A plausible mechanism is described
in Appendix A.
\qpar

In the present note we identify and study two deviations
with respect to the hyperbolic fall. What is our purpose
in so doing?
\qpar
We wish to see if the infant death rate can be used 
as a kind of spectrometer from which information can
be derived about the complex mechanisms which lead
to the regulation of the immune system.
\qpar
To carry out this program
the paper takes the following steps.
\qbu First, we must identify the time intervals
during which the system undergoes major transitions.
\qbu Once transitions have been detected one must make
sure, through comparative analysis, that they 
have a broad validity and are not just
brought about by specific local conditions.
\qbu The next step is to identify the factors
through which these deviations can be explained.
It will be seen that
the first deviation is related to viral infections
while the second is related to bacterial infections. 
There are good reasons to think that
the second deviation is due to the transition 
from protection provided
by maternal antibodies to the building up of
an endogenous immune system. In its shape this
bump is very similar to the one observed in fish during
the transition from yolk sac feeding to feeding
on exogenous sources of food. It seems that in both cases
the transition brings about a fragility which results
in the transient death rate peak that is observed.

\qI{Deviations from the hyperbolic law of infant mortality}

\qA{Identification of the time intervals of the transitions}

In Fig. 1 it can be seen that between 0 and 10 years
the infant death rate decreases by several orders of
magnitude. Without a vertical log-scale, the whole curve
would be crushed and compressed against the x-axis.
The log-scale makes
the curves more readable but Fig. 2a still does not
reveal any clear signal. In such a situation a 
common trick is to take the ratio of successive 
curves in order to filter out the huge variations.
This is done in Fig. 2b for the case of Switzerland
and the United States.

%
\begin{figure}[htb]
\centerline{\psfig{width=17cm,figure=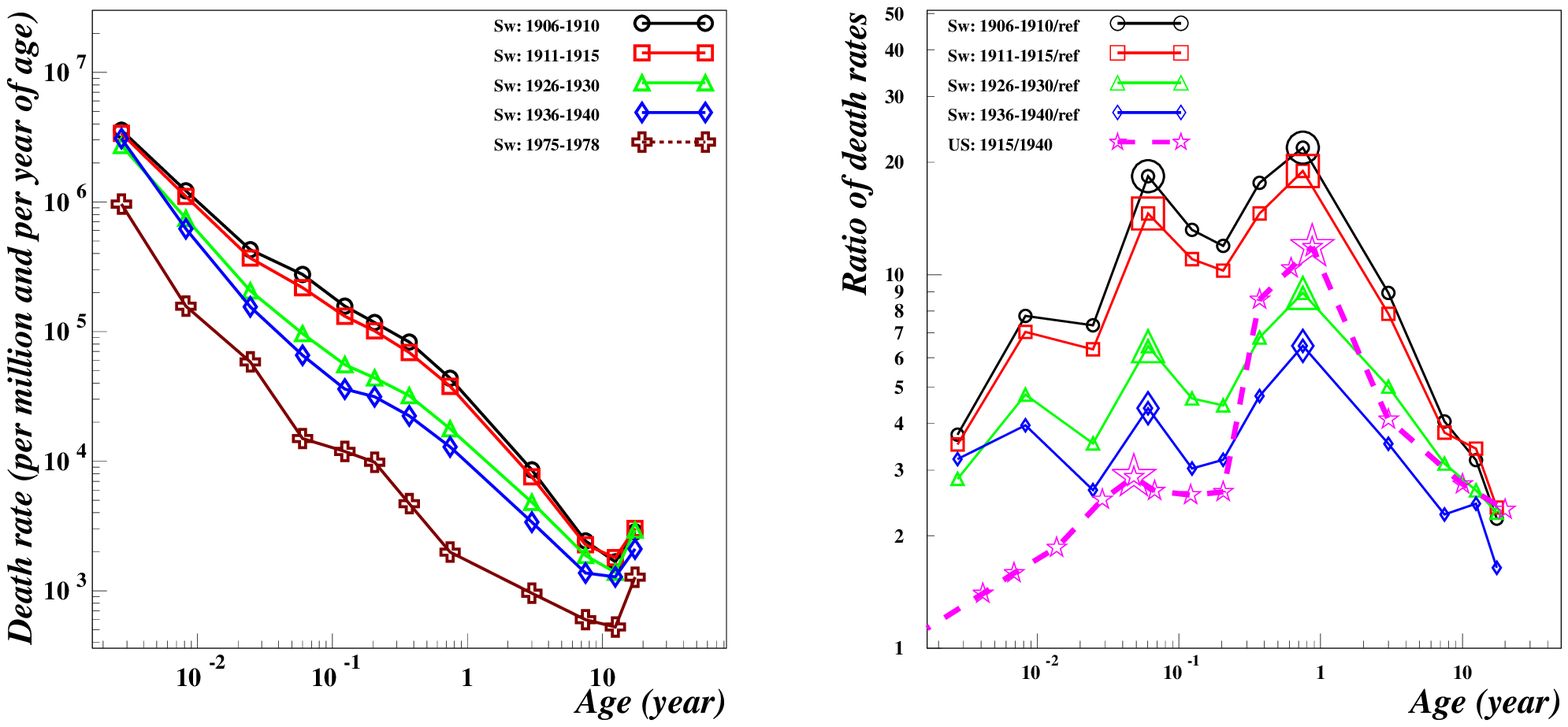}}
\qleg{Fig.\qhu 2a,b\qhv Death rates for all causes 
in Switzerland and the United States.}
{(a) Death rates in Switzerland in successive decades.
(b) Death rate ratios in Switzerland and the United States.
The ``ref'' label corresponds to the rates for 1975-1978.} 
{Sources:  Switzerland: Encyclop\'edie statistique de la Suisse.
Statistique historique. Sant\'e. Mortalit\'e et
causes de d\'ec\`es [Statistical Encyclopedia of Switzerland.
Historical Statistics; section: Health, Mortality by causes
of death.]; available on Internet at the following address:\qL
http://www.bfs.admin.ch/bfs/portal/fr/index/infothek/lexikon/lex/2.html\qL
USA: Linder and Grove (1947, p. 574).}
\end{figure}

The death ratios reveal two peaks, one at 9 days
and the other at 300 days after birth, which are common
to the two countries. In addition the curves for
Switzerland show a steady decrease 
in the amplitude of these peaks
The fact that
there are only few data points to cover the whole age interval
means that the locations of 9 and 300 days
are defined with substantial error bars which, however, are
difficult to estimate precisely.
\qpar

We can conclude this first experiment by saying
that there are two indentations on the  
curves of the age-specific mortality
from ``all causes'' which are present both in
Switzerland and in the US. 
While  not easy to identify on the rates
themselves, they become clearly visible on the ratios of
the death rates for successive decades.
The amplitudes of these indentations become larger
as one moves back in time.

\qA{Identification of the underlying diseases}

Fig. 2b suggests that the deviations 
were stronger in the first half
of the 20th century; however, they may still
exist (albeit in reduced form) in present-day data.
If this assumption is correct it will make their identification
much easier because we will be able to use the ``CDC WONDER''
database which covers the period 1999-2014
and provides unparalleled accuracy as to the causes of death. 
The fact that the database covers 16 years and
for a large country is important
because even if the effect that we wish to detect is much
smaller than in the early 20th century we may nevertheless
have a chance to see it.
\qpar

As one goes through various causes of death, e.g.
cancer, heart diseases, malformations, metabolic diseases,
enteritis, 
bacterial diseases, viral diseases, it becomes quickly
clear that only the last two in this list show visible peaks.
These are shown in Fig. 3. 

%
\begin{figure}[htb]
\centerline{\psfig{width=16cm,figure=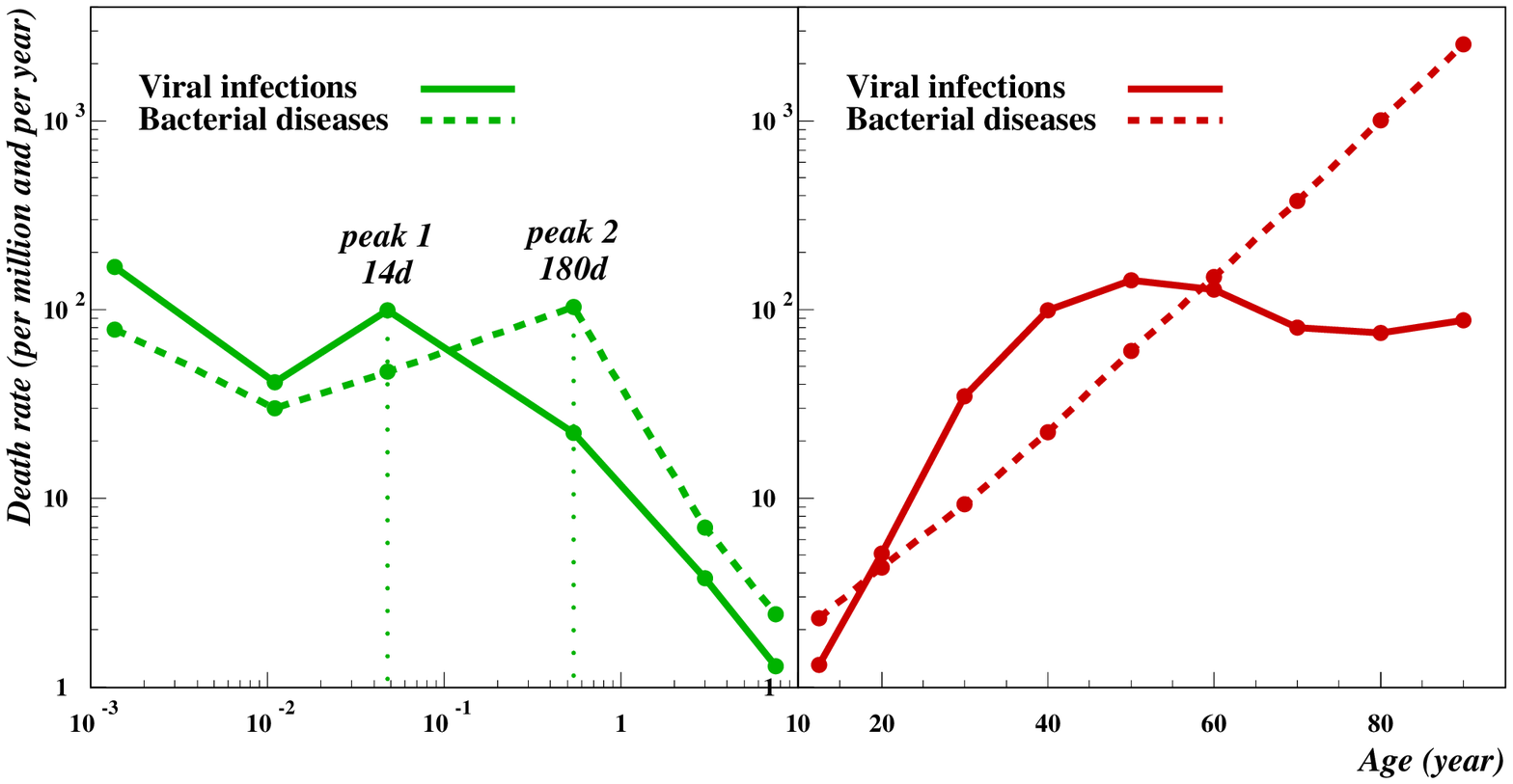}}
\qleg{Fig.\qhu 3\qhv Death rates for viral and bacterial diseases,
USA, 1999-2014.}
{The ICD10 code numbers for viral infections and bacterial 
diseases are $ (A80-B34) $ and $ (A00-09+A30-49) $ respectively.
The two peaks are found approximately
at the locations where they were expected.
The right-hand side of the graph shows that
in contrast with  bacterial diseases (as well as most other diseases)
the age-specific death rate of viral infections deviates strongly
from the Gompertz-like exponential growth. The reason for
that deviation remains an open question.}
{Source: CDC WONDER database, 1999-2014.}
\end{figure}
Can we identify them with the indentations displayed in
Fig. 2b. Two reasons make this identification plausible.
\qbu The peaks displayed for viral infections on the one hand
and bacterial diseases on the other hand are roughly
positioned where we expect them. However, this reason
is not completely compelling because of the lack of accuracy
due to the small number of data points. Over the first
year after birth the data given by WONDER are limited to
the standard postnatal intervals: 1st day, 1-7d
(early neonatal), 7-28d (neonatal), 28-365d (late neonatal).
\qbu Are the amplitudes of the peaks of Fig. 3 consistent
with what we expect?
The steady decrease of the amplitude of the 
indentations seen in Fig. 2b makes us expect
fairly small deviations in the period 1999-2014. This is
indeed the case. For the 16 years of the period under
consideration
the first three data points of the curve
for viral infections correspond
to a total number of deaths of 30, 44 and 371 respectively,
that is to say 1.8, 2.7 and 23 deaths annually. These
are very small numbers%
\qfoot{That is why we cannot give similar data for
Switzerland. As its population is 37 times smaller than 
the US population one would not see a single death.}%
.
Incidentally, we see that
more age intervals would lead to even smaller numbers
(hence more statistical fluctuations) and therefore would not
be very useful. In addition, it is clear that the
contribution of these deaths to the total mortality
is very small%
\qfoot{At the peak value of viral diseases, the 371 deaths
represent only 0.65\% of the total number of 56,616 deaths
in this age interval.}
and is therefore undetectable in the curve for ``all causes''.
\qpar

A last comment is in order regarding the left-hand graph
which covers 
the adult age interval 10-90. Strictly speaking,
this graph is not necessary for the present investigation.
However, it may be of interest because of the sharp
difference 
in shape of the two curves. In old age, the bacterial death rate 
is 200 times larger than the viral death rate.
No doubt, such a huge difference has a biological
significance.

\qI{Magnification of the signatures of viral/bacterial diseases}

In the previous sections we were able to locate the peaks
and to identify the diseases that they describe. In addition we
saw that, unsurprisingly, these peaks had been markedly reduced in
amplitude over the past century. Thanks to this understanding, we 
can now get a more accurate picture of these signatures.
For that purpose we need to introduce two improvements.
\qbu We need to view these diseases in the past when their
amplitude was still larger.
\qbu We need to find datasets which provide as many data points
as possible over the age interval $ (0,10) $.
\qpar
As we already observed, these conditions go hand in hand
in the sense that the second would be useless if the number
of deaths is too small for then most of the age intervals
would have small
numbers of deaths that would give rise to large 
statistical fluctuations.
\qpar

When one wants to go back in time one 
must be careful about two things.
\qbu There is a major difference between the data provided
by the CDC WONDER database and those given in the
digitized paper volumes of the ``Vital Statistics of the 
United States'' in the sense that the VSUS volumes give the death
numbers for a {\it selection} of causes whereas WONDER
gives them for {\it all} causes. 
\qbu In addition, the list of death causes 
(and also their definition) has been changing
with successive revisions of the ``International classification
of diseases''.
\qpar

For our present purpose, this has two direct implications.
\qee{1} In the VSUS volumes the cause ``Viral diseases'' 
appears only in 1968; thus we cannot go back in time
earlier than 1968. 
Moreover in the first years after being
introduced this entry has fairly irregular data. That is why
we considered the time interval 1973-1975.
\qee{2} The entry ``Bacterial diseases'' does not
exist in the VSUS volumes and as the entries are only
a selection it would be hazardous to set it up
by combining several sub-entries. For this reason
we prefer to focus on one specific bacterial disease, namely
tuberculosis. As is well known, tuberculosis 
was the major cause of death in the United States in the
early 20th century.

\qA{Viral diseases}

%
\begin{figure}[htb]
\centerline{\psfig{width=10cm,figure=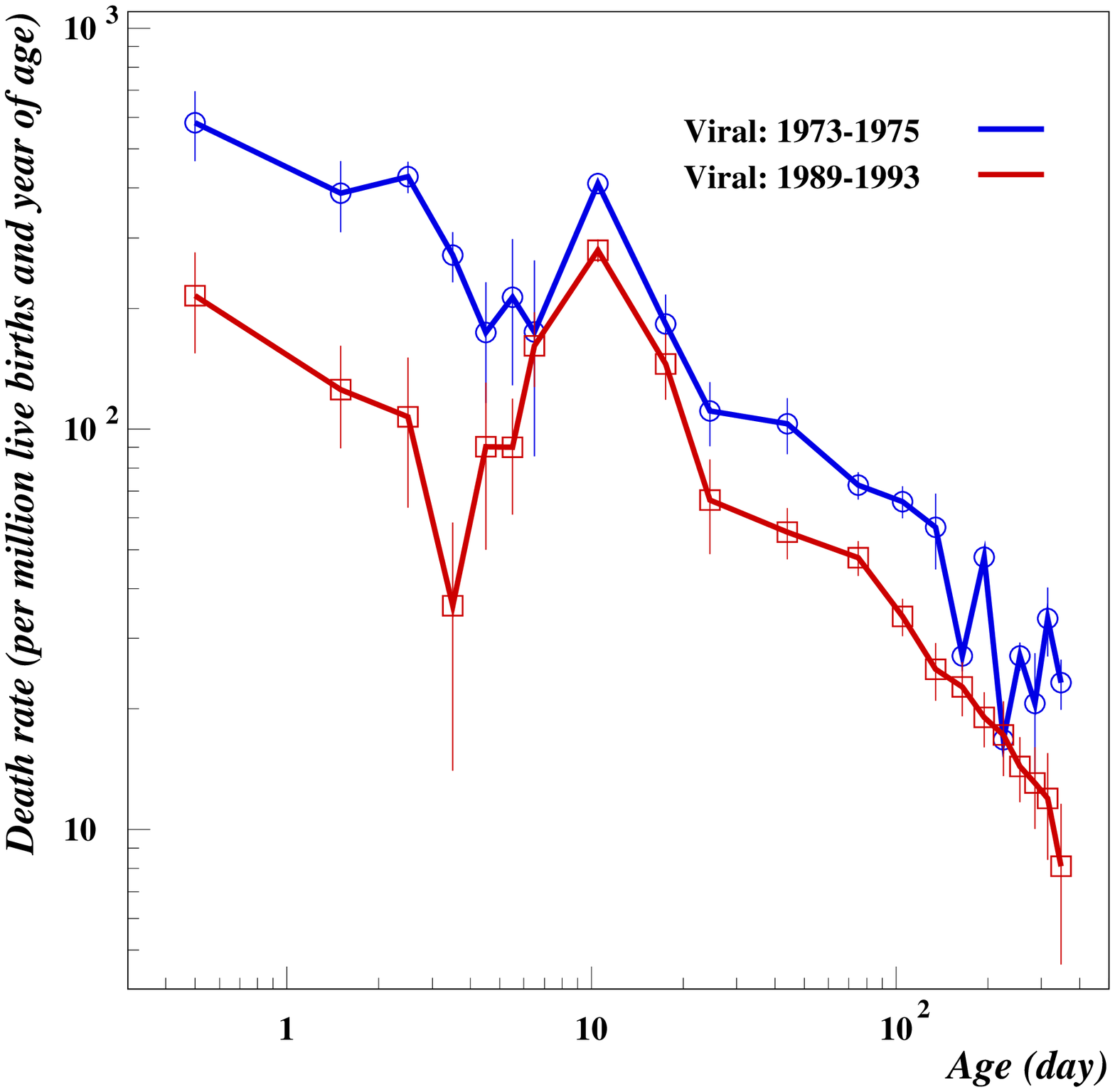}}
\qleg{Fig.\qhu 4\qhv Infant death rates for viral diseases,
USA, 1973-1975 and 1989-1993.}
{The peak value occurs at the age of 10.5 days after birth.
The error bars are $ \pm $ standard deviation.
This graph is very similar to graphs describing the 
yolk sac effect for fish (see for instance the
graph for sturgeons in Berrut et al. 2016, graph 4a).
This is not surprising because yolk is known
to have an anti-body content (Kovacs-Nolan et al. 2012)
which gives an immunity
facet to the yolk-sac effect.}
{Source: Vital Statistics of the United States
for the corresponding years.}
\end{figure}

%
\begin{figure}[htb]
\centerline{\psfig{width=10cm,figure=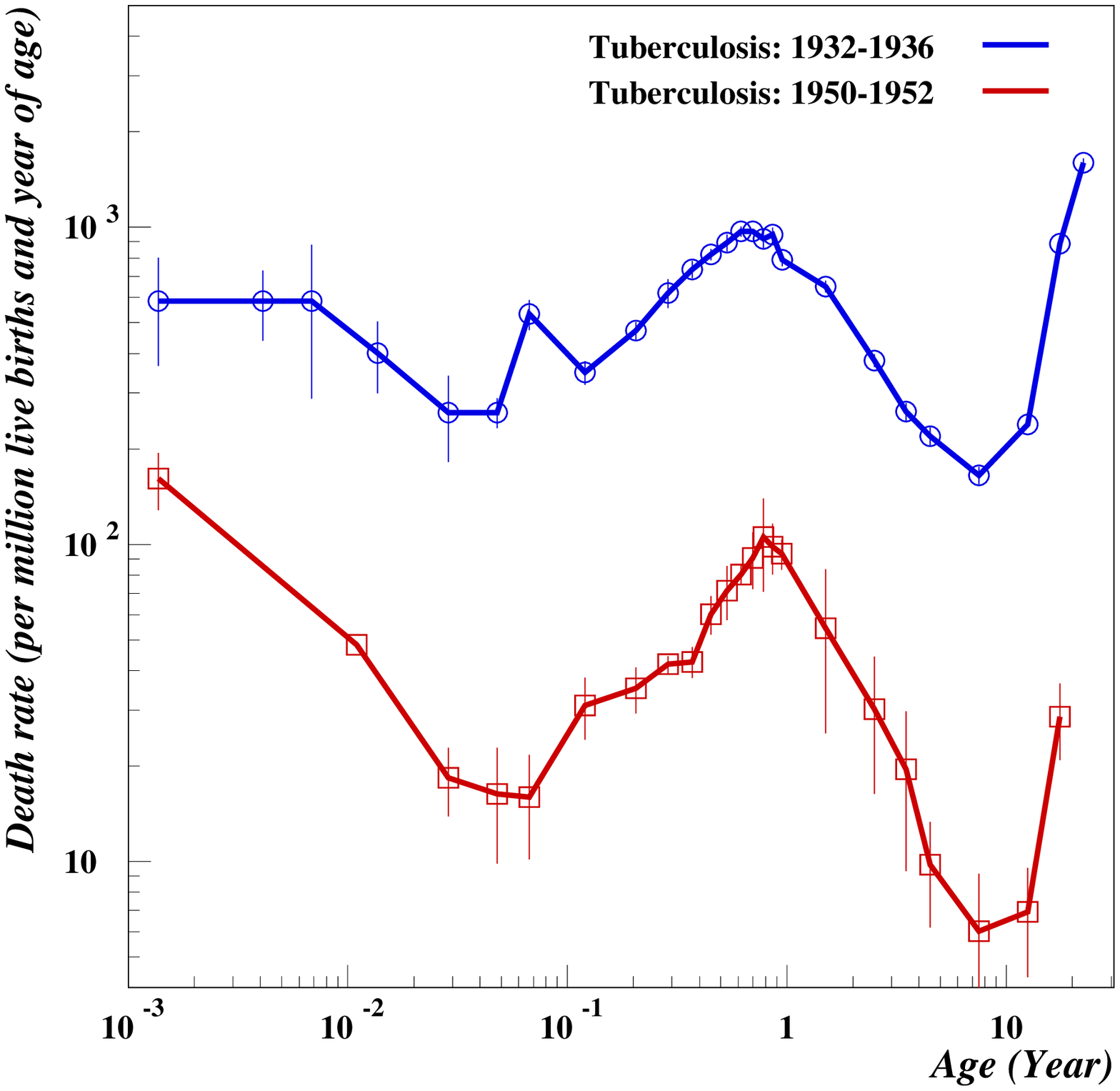}}
\qleg{Fig.\qhu 5\qhv Infant death rates for tuberculosis,
USA, 1932-1936 and 1950-1952.}
{The peak value occurs at the age of 290 days after
birth.
The error bars are $ \pm $ standard deviation.}
{Source: Vital Statistics of the United States
for the corresponding years.}
\end{figure}

Fig. 4 shows the peak of viral diseases. It extends from
5 to 22 days after birth with a peak value at 10 days.
This peak is not only present in the US for all years for which
data are available, we have also seen 
at he beginning that it exists as well in Switzerland%
\qfoot{Unfortunately the Swiss Office of Statistics 
does not seem to have data by age for this ICD entry.
Thus, one cannot make a direct comparison.
However, we are convinced that once Swiss data become
available they will display a similar peak.}%
.
What questions do these curves raise?
Just as illustrations one can mention the following points
(probably medical doctors would raise many others).
\qbu One may wonder why there was an overall decrease
 over the 16 years between the two time periods.
As far as we know, between 1973 and 1993 there was no
vaccination before the 10th day after birth.
\qbu Even if one accepts the overall reduction as a
fact, it can be observed that the reduction was
much smaller around the peak value than elsewhere;
this raises another question.

\qA{Bacterial diseases: tuberculosis} 

Fig. 5 shows the peak for tuberculosis considered as
representative
of the broader class of bacterial diseases.
It is much (130 times) wider
than the viral peak and extends from
25 days after birth to the age of 6 years.
This peak is not only present in the US 
it is also present in Switzerland. Because the disease
was very serious at the end of the 19th century fairly
detailed data are available for 1877-1881.
The corresponding curve is parallel (albeit higher
as would be expected) to the US curve in 1910.
The bacterial peak has also a larger amplitude 
than the viral peak but not in
proportion to the width ratio. The peak-to-bottom
ratio is 3 for the viral peak and 6 for the 
tuberculosis peak.

\qI{Conclusion}

\qA{Spectroscopy of age-specific infant death rates}

An alternative and more specific title for the present paper
could have been ``Spectroscopy of infant death rates''.
We decided against it because it could have been confusing
for some readers. However, the parallel with spectroscopy
conveys fairly well the idea on which this paper relies.
In spectroscopy  one analyses 
radiation intensity as a function of wavelength
in order to detect peaks which in turn will identify
emission frequencies characteristic of specific atoms. 
Here, we analyze infant death rates as a function of age
in order to characterize internal effects such as
the response 
to exogenous factors. This paper focused
on pathogens but it is also conceivable to analyze the
response to a lack of food, a sudden change in temperature
or a change in oxygen level%
\qfoot{Such a study could be done by analyzing the 
infant death rate function of populations living 
over 3,500 meters. The worldwide population living
permanently over 2,500 is estimated at 140 million.}
.
\qpar

A key point was to use the death rate ratio which
filters out what is common to two death rate functions
and thereby reveals in what respect they are different
even if this difference is a fairly small component.
Here we have used this tool to compare the death rates
in different time periods but such a
a differential comparison could also be done for two
regions or countries.
Then, in the second part, we have shown how to identify
the diseases responsible for the anomalies.
Finally,  by going back to earlier decades we could
give a fairly accurate
description of the viral and bacterial peaks. 
\qpar

As this paper is written by physicists, the question
of how to interpret
the shapes of the peaks given in Fig. 4 and 5
is left open.
The authors limited their task to
presenting the evidence as clearly as possible.
Nevertheless a few explanations regarding
the response of the immune system are given below.

\qA{Transition stages of the immune system}

The broad bacterial peak is probably in relation
with the transition from immunity based on 
maternal antibodies%
\qfoot{As a proof of the effectiveness of these
antibodies, Moreina et al. (2007) mention the
fact that babies with agammaglobulinemia (a deficiency in
the production of antibodies) are nevertheless well protected
against bacterial infection for up to 6 \hskip 0.5mm months. Then,
maternal antibodies wane over a period of 6 to 12 \hskip 0.5mm months.
Moreover,
maternal antibodies in all species have been reported to 
inhibit antibody generation after vaccination.
This makes vaccination before 6 months fairly ineffective.}
to an autonomous immune system.
Niewiesk (2014) mentions that 
the inability of the immune system of the late neonate 
to fully respond to an antigenic stimulus
has also been observed in several other mammals,
e.g. in pigs, cows, horses, mouses, rats. 
That allows comparative studies in terms of age-specific death 
rates. This will be our next objective.

\appendix

\qI{Appendix A. Plausible model for the hyperbolic fall}
 
With this plausible model our purpose is to show that it
is not difficult to imagine a mechanism which produces
an hyperbolic fall of the death rate. We call it a 
``plausible'' model because it just one possibility
among an indefinite number of possible models%
\qfoot{Three other models which also lead to an
hyperbolic fall are described in the arXiv version
of Berrut et al. (2016).}%
.
This model as well as the previous
ones are unsatisfactory in the sense 
that they do not allow any testable prediction; this is
because the parameters which they include cannot
be measured independently.
\qpar

The mechanism of the model is explained in Fig. A1.
%
\begin{figure}[htb]
\centerline{\psfig{width=16cm,figure=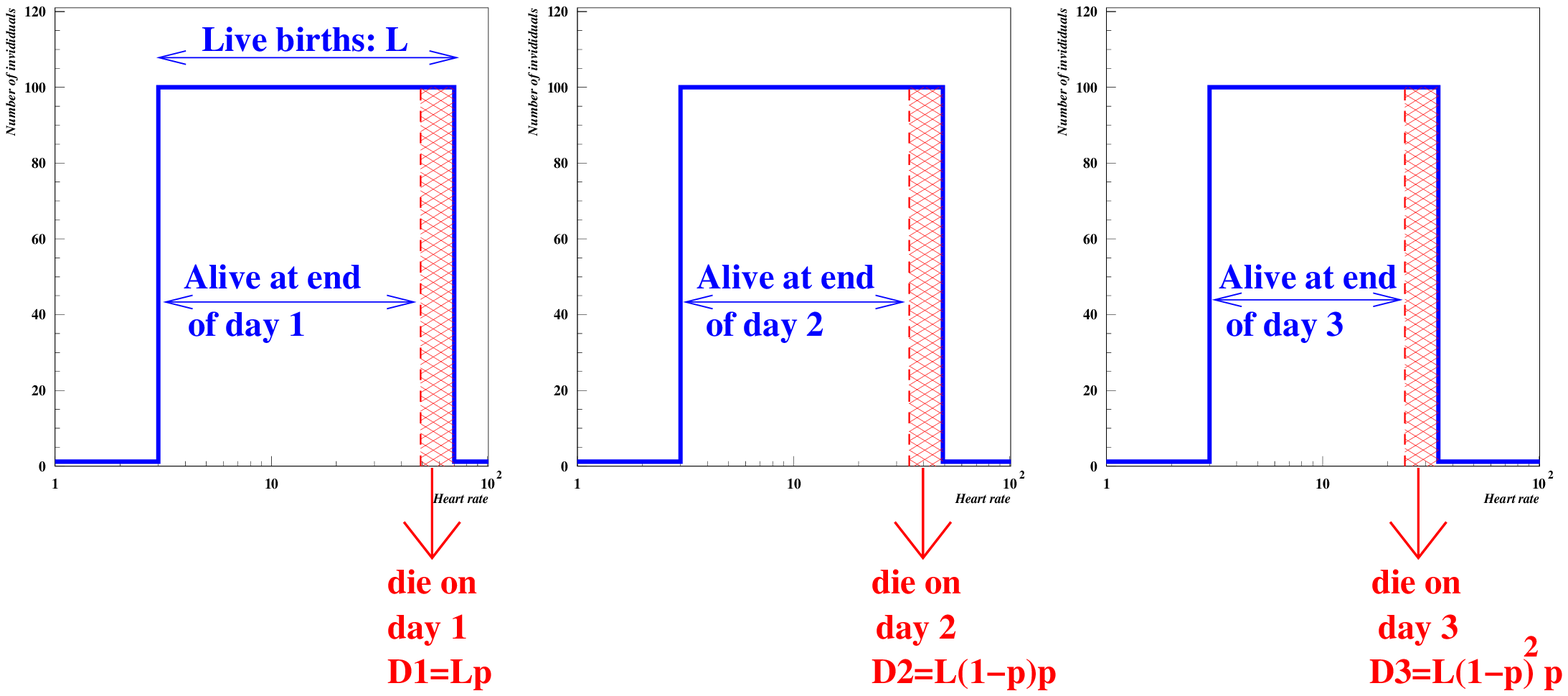}}
\qleg{Fig.\qhu A1\qhv A mechanism which leads to
an hyperbolic fall of the death rate.}
{It is because the x-scale is logarithmic that the
fraction $ p $ is represented by intervals of same lengths.}
{}
\end{figure}

We suppose that there is a spread in a physiological
variable. In Fig. A1 we considered heart rate.
It is natural to assume that those individuals for whom
the variable is farthest away from the reference
value will die on the first day (we denote them by $ D_1 $)
that those for whom
the variable deviates less will die later, say
on the second day ($ D_2 $) and so on.
It is also fairly natural to assume that the $ D_i $ are
a given fraction of the the population $ L_i $ alive at the
beginning of that day: $ D_i=L_ip $.
Thus, the successive
values of $ D_i $ become a geometric progression. 
As a result, the
infant death rate, defined as the number of deaths
divided by the number of live births, will also be a geometric
progression: $ \mu_b(t)\sim (1-p)^t $.\qL
Note that to make the model testable one should
be able to estimate $ p $ independently.

\qI{Appendix B. Experimentation in physics versus biology}

\qA{Experimentation in biodemography}

Our approach is to start from patterns and regularities observed in
humans and then to see whether (and to what extent)
they are shared by other species. Human demography belongs
to the social sciences in the sense that one cannot do experiments.
In contrast in the field usually  referred to as 
``biodemography'', it is possible 
to do experiments. One would therefore expect
experimentation in biodemography to be fairly close in its
essence and objectives
to experimentation in physics. That is not so however
and in this Appendix we try to understand the reasons and 
implications.

\qA{At what level should one observe complex systems?}

Nowadays biologists have at their disposal
highly sophisticated techniques of
investigation which give them access to detailed
mechanisms at the level of proteins and other macromolecules.
While such techniques represent a great opportunity, there
is also a possible downside in the sense that to have 
too many details
may hinder global understanding. 
When a description of apoptosis
(i.e. programmed cell death) involves dozens of 
physical processes (e.g. folding, unfolding, sticking, diffusing)
and chemical reactions
it becomes difficult for a human mind to make sense of it.
To say it differently, if in their exploration of
matter physicists of the eighteenth and nineteenth century
have had at their disposal Raman spectroscopy,
slow neutron scattering and X-ray diffraction they may 
have refrained from
proposing some of the crude, yet nonetheless very useful,
phenomenological
laws on which the physics of that time relied. It was
a piece of good luck that in the development of physics
progress in our understanding and means of observation
progressed in sink.

\qA{The key of comparative analysis}

However, if the availability of precise means 
of observation at the 
molecular level is one part of the explanation, it cannot
in itself explain why cross-species comparative studies 
have become very rare. 
In physics, the comparative perspective
was essential from the very beginning.
Once a new physical effect 
had been identified the main goal was always to determine 
its scope of occurrence. For instance,
gravitation coupled with Newton's law describes
the fall of apples, but also the phenomenon of the
tides, the movement
of the Moon around the Earth and (at least to some extent) the
rotation of galaxies around their center.
\qpar

When we started our investigation of
infant mortality we were surprised to see
that despite of the very remarkable law governing
human infant mortality, there had been no
studies which addressed the question of whether
other species followed the same law.
Thus, our 
first investigation (Berrut et al. 2016) was to answer
this question. We found the same pattern in mammals and fish.
In our paper we used our own data for primates
and for other species we relied on existing data.
However, all such data
had been collected for reasons which had nothing to
do with the purpose that we had in mind. They were
recorded mostly for economic reasons
because for farmers it is obviously 
of interest to know postnatal mortality of
farm animals. The same observation applies to 
fish with respect to aquafarming.
Whether beyond mammals and fish the same pattern
also extends to ``simpler''
organisms (e.g. C. elegans) is at
the present moment still an open question;
it is under investigation
in a series of experiments.

\qA{Why is there a reproducibility crisis in biology 
and soft matter physics?}

Apart from the lack of comparative perspective, another
important feature of biological experiments is the fact that 
they are rarely repeated. 
A recently published editorial in the journal ``Nature''
(Vol. 533, 437, 26 May 2016) mentions that 
two-thirds of the researchers who responded to a survey
think that the current lack of reproducibility
is a major problem. 
To the mostly circumstantial reasons that are given
in the editorial one
can add a more fundamental factor which is precisely the
lack of comparative perspective already mentioned.
Indeed, if a mechanism has a broad range of validity
it makes sense to measure it with upmost accuracy.
In physics, whenever a measurement
is ``repeated'' this is almost always done for improving its 
accuracy. Actually, even in physics
an experiment is never strictly ``repeated'';
it is done under conditions 
in which one expects indeed the same effect to be observed
but these conditions are never exactly the same as
in the observations conducted by other teams.
While improving the accuracy is a major 
incentive for ``repeating'' an experiment, when
this incentive does not exist it is quite
understandable that few experiments are repeated%
\qfoot{Actually, there is currently the same concern in soft
matter physics. Why should one try to improve
the accuracy of measurements which anyway
have a very narrow range of validity? Whereas
the speed of light in vacuum is of fundamental
importance, the incentive for measuring 
with high accuracy the 
transversal speed of sound in egg yolk 
is not quite as high.}%
.

\qA{Medicine-oriented funding}
 
In the physical sciences there is a clear separation between
fundamental physics and engineering. 
The funding
of fundamental physics is largely independent of any 
perspective of practical applications. As an illustration
one can mention the funding of the ``Large Hadron Collider''
whose main success so far has been the discovery
of the Higgs boson, a particle which has a lifetime
of only $ 10^{-22} $ s. In the same line of thought
one can mention the funding of the expensive telescopes
which are necessary in astrophysics.
\qpar

On the contrary,
in biology there is no clear separation between
fundamental biology and medicine in the sense that
funding is much influenced
by the war against various diseases, e.g.
cancer, Alzheimer's disease or 
other degenerative diseases. Even for research 
of a fairly fundamental nature, discovering new treatments
is a permanent temptation for researchers because
of the fame and social rewards attached to such discoveries.

\vskip 5mm

{\bf References}

\qparr
Berrut (S.), Pouillard (V.), Richmond (P.), Roehner (B.M.) 2016:
Deciphering infant mortality. 
Physica A 463, 400-426.

\qparr
CDC WONDER is a mortality database set up and maintained
by the ``Centers for Disease Control''. It is available
at the following address:\qL
http://wonder.cdc.gov/ucd-icd10.html.

\qparr
Encyclop\'edie statistique de la Suisse.
Statistique historique. Section: Sant\'e. Sous-section:
Mortalit\'e et
causes de d\'ec\`es [Statistical Encyclopedia of Switzerland.
Historical Statistics. Section: Health. Subsection:
Mortality by causes of death.]  
Available on Internet at the following address:\qL
http://www.bfs.admin.ch/bfs/portal/fr/index/infothek/lexikon/lex/2.html

\qparr
Kovacks-Nolan (J.), Mine (Y.) 2012: 
Egg yolk antibodies for passive immunity.
Annual Review of Food Science and Technology, 3,163-182.

\qparr
Linder (F.E.), Grove (R.D.) 1947: Vital statistics rates in the
United States 1900-1940. US Government Printing Office,
Washington DC.

\qparr
Moreina (B.), Blomqvist (G.), Hu (K.) 2007:
Immune responsiveness in the neonatal period. 
Journal of Comparative Pathology 137,S27

\qparr
Niewiesk (S.) 2014: 
Maternal antibodies: clinical significance, mechanism of interference
with immune responses, and possible vaccination strategies.
Frontiers in  Immunology 5,446.

\end{document}